\documentclass[conference]{IEEEtran}
\IEEEoverridecommandlockouts

\usepackage{cite}
\usepackage{amsmath,amssymb,amsfonts}
\usepackage{amssymb}
\usepackage{algpseudocode}
\usepackage{amsfonts}
\usepackage{graphicx}
\usepackage{fancyhdr}  %
\usepackage{cases}
\usepackage{textcomp}
\usepackage{extarrows}
\usepackage{orcidlink}

\usepackage{multirow}
\usepackage{caption}
\usepackage{float} 
\usepackage{subcaption}
\usepackage{makecell}

\usepackage{booktabs}
\usepackage{array}
\usepackage{footnote}
\makesavenoteenv{tabular}

\usepackage{algorithm,algpseudocode}
\algnewcommand{\Inputs}[1]{%
  \State \textbf{Inputs:}
  \Statex \hspace*{\algorithmicindent}\parbox[t]{.8\linewidth}{\raggedright #1}
}
\algnewcommand{\Initialize}[1]{%
  \State \textbf{Initialization:}
  \Statex \hspace*{\algorithmicindent}\parbox[t]{.8\linewidth}{\raggedright #1}
}

\usepackage{multirow,tabularx}
\usepackage{multicol,mwe,float,subcaption}
\usepackage{mathtools}
\usepackage{xcolor}
\usepackage[english]{babel}
\usepackage{amsthm}
\usepackage[numbers,sort&compress]{natbib}

\makeatletter
\renewcommand{\maketag@@@}[1]{\hbox{\m@th\normalsize\normalfont#1}}%
\makeatother

\def\BibTeX{{\rm B\kern-.05em{\sc i\kern-.025em b}\kern-.08em
    T\kern-.1667em\lower.7ex\hbox{E}\kern-.125emX}}
\begin{document}

\title{Multi-Sources Information Fusion Learning for Multi-Points NLOS Localization}

\author{
	\IEEEauthorblockN{Bohao Wang\IEEEauthorrefmark{1}, Fenghao Zhu\IEEEauthorrefmark{1}, Mengbing Liu\IEEEauthorrefmark{4}, Chongwen Huang\IEEEauthorrefmark{1}, Qianqian Yang\IEEEauthorrefmark{1},}
	\IEEEauthorblockN{Ahmed Alhammadi\IEEEauthorrefmark{3}, Zhaoyang Zhang\IEEEauthorrefmark{1}, and M\'{e}rouane~Debbah\IEEEauthorrefmark{5}\IEEEauthorrefmark{6}, \IEEEmembership{Fellow,~IEEE}}
	
	\IEEEauthorblockA{\IEEEauthorrefmark{1} College of Information Science and Electronic Engineering, Zhejiang University, 310027, Hangzhou, China}
	\IEEEauthorblockA{\IEEEauthorrefmark{3} Technology Innovation Institute, 9639 Masdar City, Abu Dhabi, UAE}
	\IEEEauthorblockA{\IEEEauthorrefmark{4} School of Electrical and Electronics Engineering, Nanyang Technological University, 639798, Singapore}
	\IEEEauthorblockA{\IEEEauthorrefmark{5} KU 6G Research Center, Khalifa University of Science and Technology, P O Box 127788, Abu Dhabi, UAE}
	\IEEEauthorblockA{\IEEEauthorrefmark{6} CentraleSupelec, University Paris-Saclay, 91192 Gif-sur-Yvette, France}
	
    }

\maketitle

\begin{abstract}

Accurate localization of mobile terminals is crucial for integrated sensing and communication systems. 
Existing fingerprint localization methods, which deduce coordinates from channel information in pre-defined rectangular areas, struggle with the heterogeneous fingerprint distribution inherent in non-line-of-sight (NLOS) scenarios. 
To address the problem, we introduce a novel multi-source information fusion learning framework referred to as the Autosync Multi-Domain NLOS Localization (AMDNLoc). 
Specifically, AMDNLoc employs a two-stage matched filter fused with a target tracking algorithm and iterative centroid-based clustering to automatically and irregularly segment NLOS regions, ensuring uniform fingerprint distribution within channel state information across frequency, power, and time-delay domains. 
Additionally, the framework utilizes a segment-specific linear classifier array, coupled with deep residual network-based feature extraction and fusion, to establish the correlation function between fingerprint features and coordinates within these regions. 
Simulation results demonstrate that AMDNLoc significantly enhances localization accuracy by over 55\% compared with traditional convolutional neural network on the wireless artificial intelligence research dataset.

\end{abstract}

\begin{IEEEkeywords}
Multi-sources, information fusion, fingerprint localization, inverse, heterogeneity, regional covariant.
\end{IEEEkeywords}

\section{Introduction}\label{sec:intro}
As a key usage scenario for sixth-generation communication, integrated sensing and communication requires a low-latency and high-precision mobile terminal (MT) localization, especially in fields like the smart city, internet of vehicles, telemedicine and so on. 
Among various methods, fingerprint-based localization, leveraging unique multi-path features at each location to infer specific positions, stands out in massive multiple-input multiple-output orthogonal frequency division multiplexing (MIMO-OFDM) systems.

In particular, it treats channel state information (CSI) of multi-path features as fingerprints, and matches real-time measurements with stored fingerprints to estimate locations. 
However, this process, especially in outdoor multi-point non-line-of-sight (NLOS) scenarios, is fraught with complexities. The multi-path features that are influenced by diverse urban elements like buildings and scattered objects lead to substantial heterogeneity in fingerprint distribution. This results in an uneven regional on the relationship between fingerprints and locations, complicating the inverse problem of location inference from fingerprints \cite{xiao2018learning}.

To tackle the issue, traditional solutions involve dividing a large area into grid cells, assigning labels within each cell, and iteratively refining the search until the closest fingerprint in the dataset is matched \cite{bittencourt2018proposal, vo2015survey, sun2018single, peng2019decentralized, sun2019fingerprint, li2020deep, del2019localization, gante2020deep, gong2023transformer}.
Although manual division of areas is easy to operate, the localization accuracy can be sensitive to the specifics of regional divisions. Furthermore, due to the heterogeneity issue, while fingerprints of adjacent positions are more likely to exhibit high similarity, existing works do not guarantee that all fingerprints within a cell share a high similarity, no matter the cell is regular or not. 

Building upon the manually divided areas, several artificial intelligent (AI) techniques have been proposed to build the inverse correlation function between fingerprints and locations \cite{peng2019decentralized, sun2019fingerprint, li2020deep, del2019localization, gante2020deep, gong2023transformer, ferrand2020dnn}.
However, these approaches refine the search area hierarchically which means that there are amounts of parameters to be updated and once the coarse position of the fingerprint is predicted incorrectly, errors will accumulate to the final results.
What's more, though each location's multi-path features are unique, these approaches may fail to fully encapsulate this uniqueness by only choosing one of the received signal strength (RSS), channel impulse response (CIR), channel frequency response (CFR), and the angle-delay channel amplitude matrix (ADCAM) as the fingerprint. Experiments suggest that there exists a proportion of remote points sharing very similar fingerprints within a scene in different dataset, thereby violating the foundational assumption of independent and identically distributed (i.i.d) data and undermining the predicted results \cite{yang2013rssi, bhattacherjee2020localization}.
\begin{figure*}\vspace{-0mm}
	\begin{center}
		\centerline{\includegraphics[width=1\textwidth]{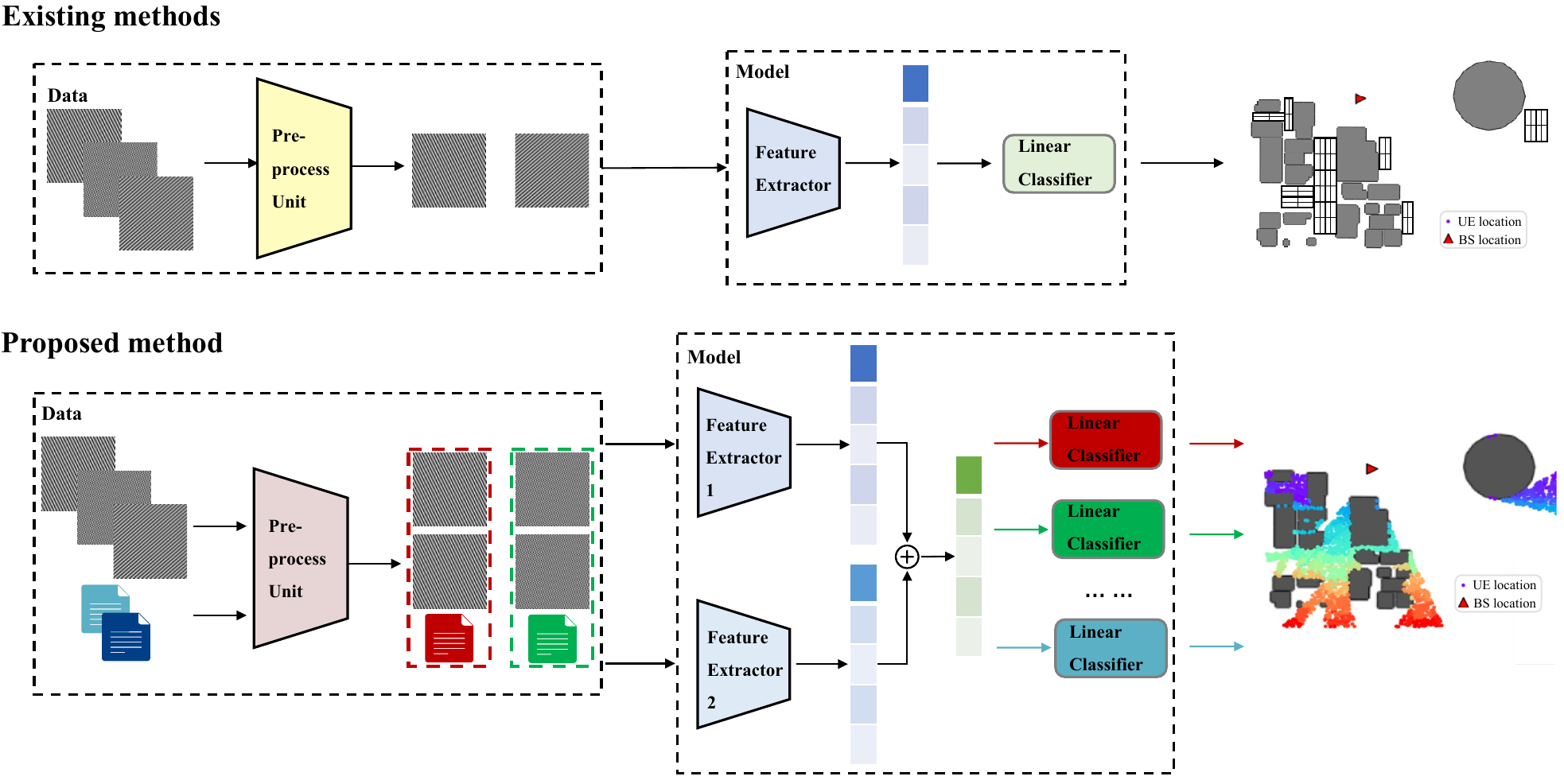}}  \vspace{-0mm}
		\caption{System Diagram of proposed AMDNLoc framework compared with existing methods.}
		\label{fig:framework} \vspace{-10mm}
	\end{center}
\end{figure*}

To solve above-mentioned issues, we propose a novel multi-source information fusion learning framework for large scale outdoor multi-points NLOS localization called the autosync multi-domain NLOS
localization (AMDNLoc) framework, which automatically and irregularly generates the classification regions based on the fusion of frequency, power and time-delay information of CSI for the first time. 
Specifically, we develop a two-stage matched filter to identify and extract all parallel feature intervals related to CFR distribution, termed parallel feature CFR (PFCFR), and classify all CFR images according to the closest PFCFR by target tracking algorithm. Then, we utilize the iterative centroid-based clustering to partition the ADCAM considering power, angle, and delay characteristics, and combine these two classification areas maintaining the same fingerprint distribution characteristics.
Then, special samples are eliminated by data cleaning and a segment-specific linear classifier array coupled with deep residual network-based feature extraction and fusion is introduced to inversely map the fingerprints to the coordinates in the respective regions while maintaining the global features. 
State of the art (SOTA) results are achieved in the ray-tracing wireless AI research dataset (WAIR-D) \cite{fenghao1, huangfu2022wair}. Also the visualization of classification areas and comparison of multiple processing are validated to ensure the enhanced adaptability and scalability of AMDNLoc.
The code is avalable in the \cite{gitAMDNloc}.



\section{System and Channel Model}\label{sec:sys}

In this section, we begin by presenting the channel model and subsequently formulate the heterogeneity problem in NLOS fingerprint localization.

\begin{figure}[ht]
	\begin{center}
		\centerline{\includegraphics[width=0.5\textwidth]{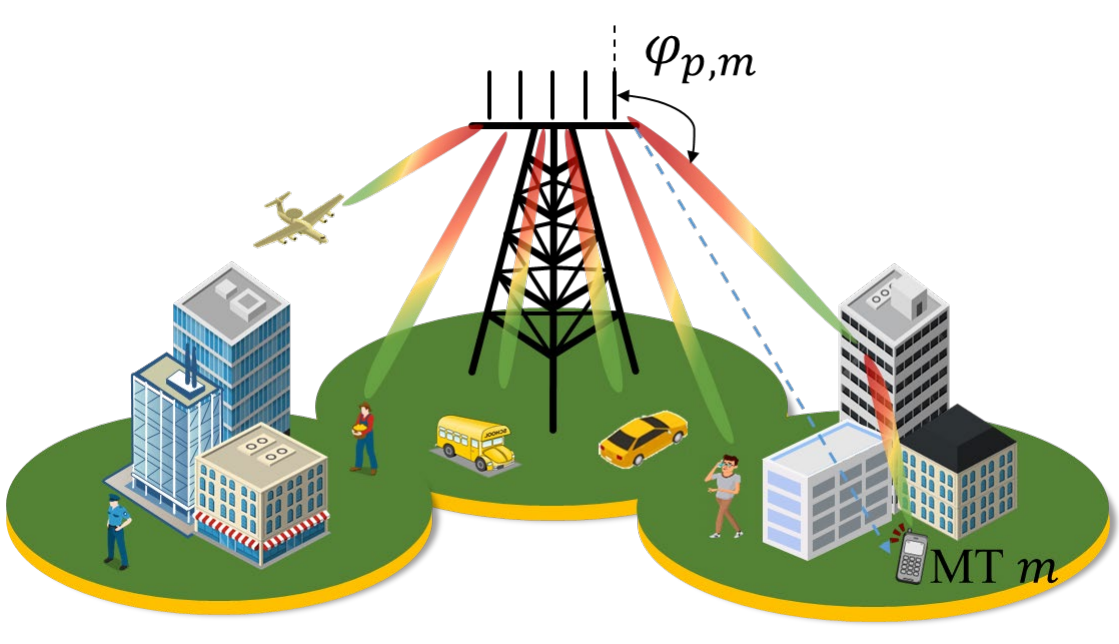}}
		\caption{ The wireless channel from MT $m$ to the BS.}
		\label{fig:system}
		\vspace{-11mm}
	\end{center}
\end{figure}

The channel transmission diagram in a MIMO-OFDM system is illustrated in Fig. \ref{fig:system}. The BS is equipped with a uniform linear array (ULA) comprising $N_{t}$ antennas. We have $m\in\{1, ..., M\}$ MTs, each equipped with a single omni-directional antenna. The angle of arrival (AOA) and the physical distance between the transmit antenna and the first receive antenna associated with the $p$-th path are denoted by $\phi_{p,m}\in(0,\pi)$ and $d_{p,m}$, respectively.

The CIR vector associated with the $p$-th path of the $m$-th user is given by:
\begin{equation}
{{\bf q}}_{p,m} = a_{p,m}{\bf e}\left(\phi_{p,m}\right),
\end{equation}
where $a_{p,m}\sim\mathcal CN(0,\sigma_{p,m})$ represents the complex gain associated with the $p$-th path, and ${\bf e}\left(\phi\right)$ is the array response vector corresponding to the AOA $\phi$. It takes the form:
\begin{equation}
{{\bf e}}\left(\phi \right) = \left[1,{e^{ - {\bar{\jmath }}2\pi { {{d\cos \left(\phi \right)} \over {{\lambda _c}}}}}}, \ldots,{e^{ - {\bar{\jmath }}2\pi { {{\left({{N_t} - 1} \right)d\cos \left(\phi \right)} \over {{\lambda _c}}}}}} \right]^T,
\end{equation}
where $j=\sqrt{-1}$, $d$ is the antenna spacing (typically $\lambda/2$ in recent MIMO systems); $\tau_{p,m}=n_{p,m}T_s$ represents the distinguishable propagation delay associated with the $p$-th path, $n_{p,m}$ refers to the sampled delay for the $p$-th path, and $T_s$ denotes the sample interval. The CFR is expressed as a sum of time-domain CIRs with varying delays:
\begin{equation} {{{\bf h}}_{m,l}} = \sum \limits _{p = 1}^P {{a_{p,m}}{{\bf e}}\left({{\phi _{p,m}}} \right){e^{ - {\bar{\jmath }}2\pi { {{l{n_{p,m}}} \over {{N_c}}}}}}},  \end{equation}
where $N_c$ is the number of subcarriers in the OFDM system, $l\in \{1,...,N_c\}$ . Then, the overall CFR matrix known to the BS can be denoted as the stack of ${{{\bf h}}_{m,l}}$, i.e.,
\begin{equation} \label{H}{{{\bf H}}_m} = \left[ {{{{\bf h}}_{m,0}},{{{\bf h}}_{m,1}}, \ldots,{{{\bf h}}_{m,{N_c} - 1}}} \right], \end{equation}

Then, authors in \cite{sun2018single} established a mapping from the CFR matrix to a sparse structure by employing DFT operations, referred to as the angle-delay channel response matrix. It is a linear transformation of the CSI computed by multiplying it with two DFT matrices.

Let us define the DFT matrix ${\mathbf{V}}\in \mathbb{C}^{N_t\times N_t}$ as:
\begin{equation*}
{[{\mathbf{V}}]_{z,q}} \triangleq \frac{1}{{\sqrt {{N_t}} }}{e^{ - j2\pi \frac{{\left( {z\left( {q - \frac{{{N_t}}}{2}} \right)} \right)}}{{{N_t}}}}},
\end{equation*}
and ${\mathbf{F}}\in \mathbb{C}^{N_c\times N_c}$ as:
\begin{equation*}
{[{\mathbf{F}}]_{z,q}} \triangleq \frac{1}{{\sqrt {{N_c}} }}{e^{ - j2\pi \frac{{zq}}{{{N_c}}}}},
\end{equation*}
Therefore, the ADCAM of user $m$ is defined as:
\begin{equation}
{\left[ {{\bf {A}_m}} \right]_{z,q}} = \mathbb {E}\left\lbrace {\left| {{\left[{{{\mathbf{V}}^H}{\mathbf{H}_m\mathbf{F}}}\right]_{z,q}}} \right|} \right\rbrace,
\end{equation}
where the $(z,q)$ element of the ADCAM represents the absolute gain of the $z$-th AOA and $q$-th delay of the channel.

As we know, fingerprint is determined by the locations and a specific configuration of the propagation environment, given by:
\begin{equation}
\textit{Fingerprint} = \boldsymbol{\mathcal F}(\boldsymbol{p}, pv),
\end{equation}
where $\textit{Fingerprint}$ represents the channel characteristics used for localization. $\boldsymbol{p} := [x, y]$ denotes the 2D positional coordinates of the users, and $pv$ encompasses the system parameters which remain constant during training and testing.

Traditional fingerprint localization aims to inversely map the $\textit{Fingerprint}$ back to $\boldsymbol{p}$. However, the impact of different buildings on various areas differs, so that changes in the fingerprint in different areas play different roles in affecting user locations. This leads to the issue of fingerprint heterogeneity in distribution.
Therefore, the inverse process of fingerprint extraction should be expressed as:
\begin{equation}
\hat{\boldsymbol{p}}_{\text{real}} = \boldsymbol{\mathcal F^{-1}}(\textit{Fingerprint}, pv, cv),
\end{equation}
where $cv$ represents a covariate related to the region, weather, and other factors. Our objective in this research is to eliminate the regional covariate from fingerprint localization and establish a one-to-one relationship solely between $\boldsymbol{p}$ and $\textit{Fingerprint}$.

\section{AMDNLoc Multi-sources Framework}\label{sec:sys}
In this section, we delve into the heterogeneity matched filter, the design of the network structure, and the detailed network training methodology within the AMDNLoc framework.

\subsection{Heterogeneity Matched Filter}
In tackling the complex issue of heterogeneous fingerprint distribution, our research has led to the development of a specialized classification approach to uniquely addresses the distinct distribution characteristics inherent to both CFR and ADCAM.

\subsubsection{PFCFR}
\begin{figure}[htbp]
	\centering
	\begin{subfigure}{0.325\linewidth}
		\centering
		\includegraphics[width=0.9\linewidth]{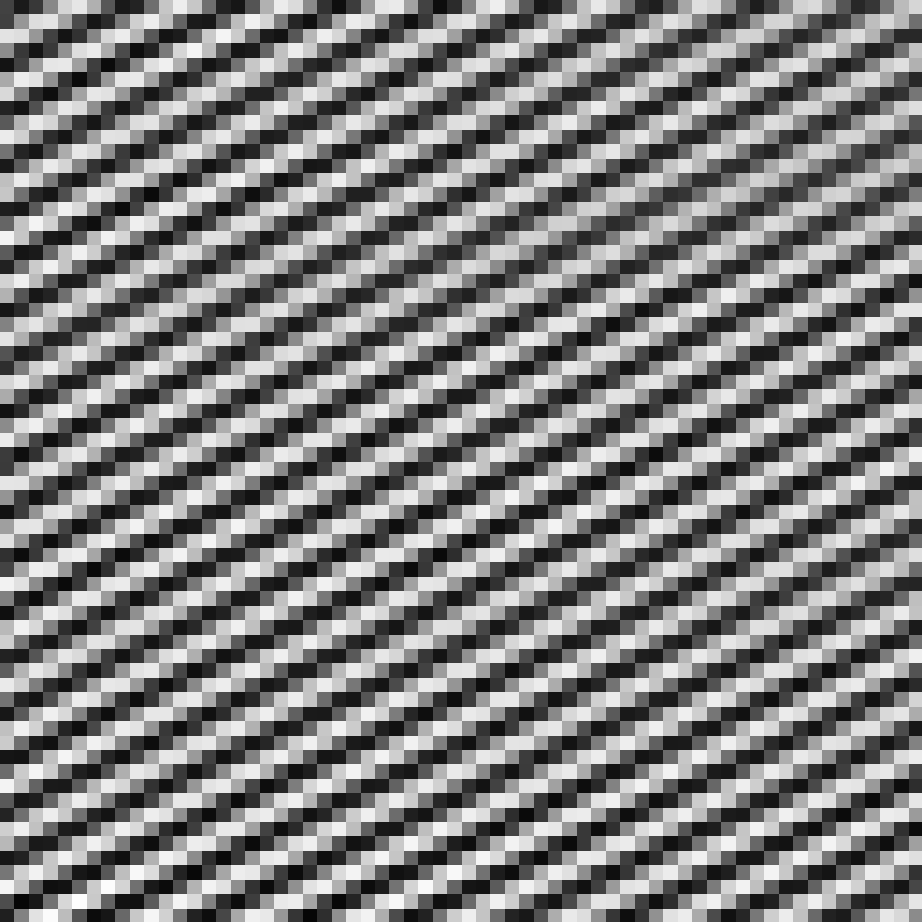}
		\caption{ (101.22 49.29)}
		\label{chutian3}
	\end{subfigure}
	\centering
	\begin{subfigure}{0.325\linewidth}
		\centering
		\includegraphics[width=0.9\linewidth]{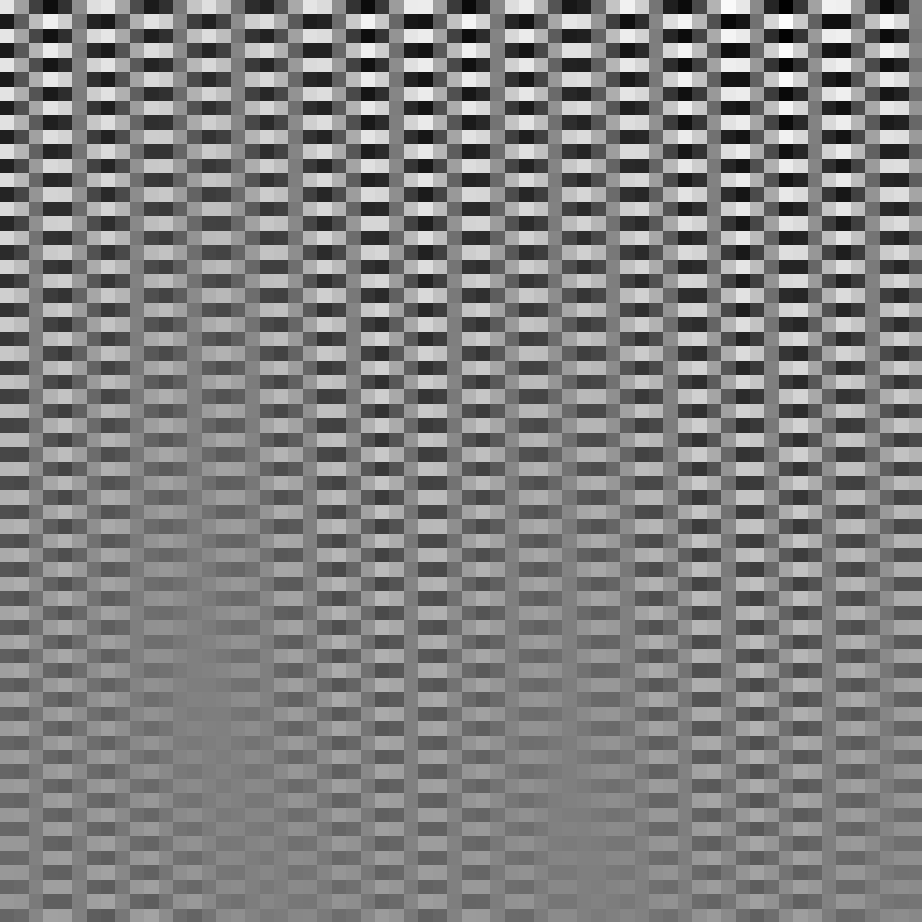}
		\caption{ (73.85 30.77)}
		\label{chutian3}
	\end{subfigure}
	\centering
	\begin{subfigure}{0.325\linewidth}
		\centering
		\includegraphics[width=0.9\linewidth]{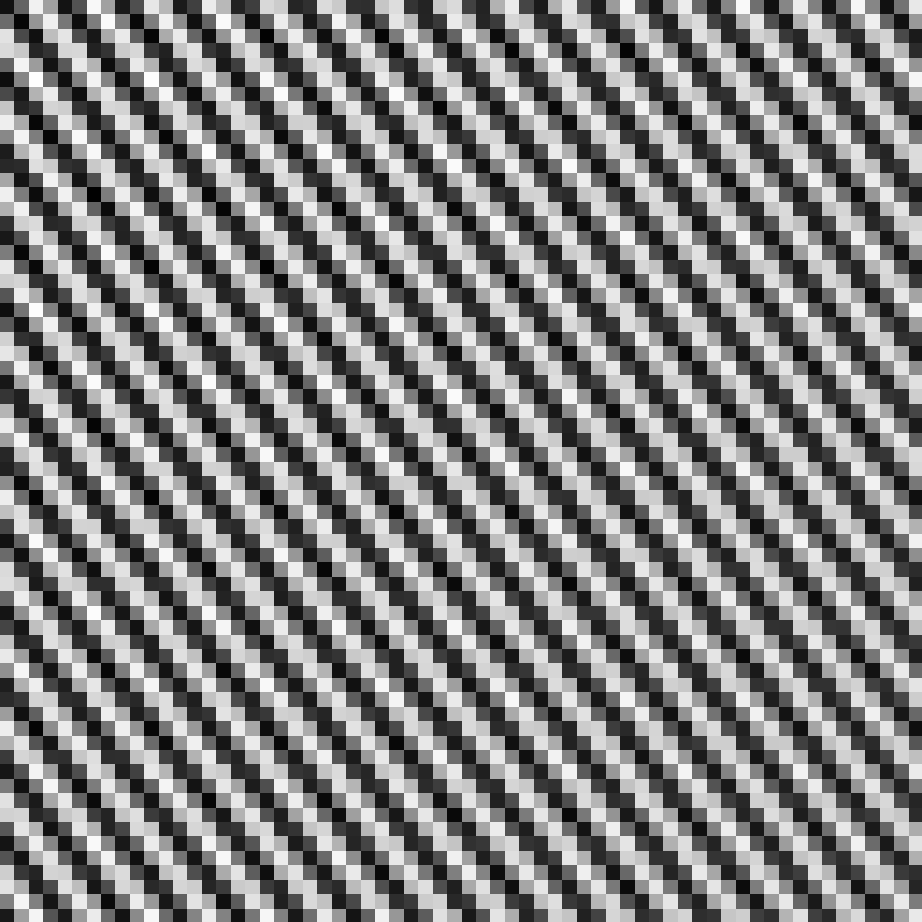}
		\caption{ (144.64 93.75)}
		\label{chutian3}
	\end{subfigure}
	\caption{CFR example figure of randomly selected MTs in the 00743 scenario of WAIR-D}
	\label{fig:CFR_example}
	\vspace{-6mm}
\end{figure}
We commence by depicting the matrix $H$ from Eq. \ref{H} as a two-channel grayscale image, as illustrated in Fig. \ref{fig:CFR_example}. In this representation, the $x$-axis corresponds to the carrier frequency, while the $y$-axis denotes the antenna pair. A consistent observation across the dataset reveals the presence of distinct horizontal and vertical translational structures within certain regions in CFR, which we refer to as PFCFR. 
PFCFR is a key indicator of data distribution of CFR. This distribution is largely shaped by physical phenomena such as refraction, reflection, and diffraction, which exhibit more pronounced similarities in adjacent positions. Consequently, it is a logical inference that CFR sharing similar PFCFR characteristics are likely to be located in geographically close areas. This proximity-based similarity allows for an effective categorization of these CFR according to their respective PFCFR values, providing valuable insights into the spatial dynamics of data distribution influenced by channel properties.

To leverage this insight, we propose a novel two-stage pipeline matched filter outlined in Alg. \ref{alg:PFCFR1} and Alg. \ref{alg:PFCFR2}. Inspired by the target tracking algorithm, the filter is designed firstly to extract PFCFR from all samples as the template and CFR as the source image for CFR categorization. 
To mitigate the risk of coincidental similarities in specific regions between two images, our approach incorporates dual template regions. We define $T_1(i,j,x',y')$ and $T_2(i,j,x',y')$ to represent the pixel values from the upper-left and lower-right corners, respectively, of the $j$-th source image. Here, the notation $(i,x',y')$ specifies the pixel coordinates in the $i$-th template image.
The application of this filter involves systematically shifting the origin of the template image across each point in the source image. The similarity between the template and source images is calculated by aggregating the products of their corresponding pixel values across the entire span of the template. To quantify this similarity, we employ the normalized cross-correlation, denoted as ${\mathbf{E_n(i,j)}}$, given by:

\begin{scriptsize}
\begin{equation}
\hspace{-1.5mm}
    \begin{aligned}
        {\mathbf{E_n(i,j)}}\!&=\! \max\! \frac{\sum_{x',y'}\!{T_n(i,j,x',y')I(j,x\!+\!x',y\!+\!y')}}{\sqrt{\sum_{x',y'}\!{T_n(i,j,x',y')^2} \sum_{x',y'}\!{I(j,x\!+\!x',y\!+\!y')^2}}}{,} 
    \end{aligned}\label{eq:10}\end{equation}
\end{scriptsize}  
where $n={1,2}$, ${\mathbf{E_n(i,j)}}$ is between 0 and 1, and the closer it is to 1, the higher the similarity. Consequently, the process of our proposed two-stage pipeline matched filter can be summarized as follows:

\textbf{Initialization:} 
We set $\tau _c$ to a value greater than $M$, which serves as a threshold for determining whether the $j$-th image is matched. Then, $\tau _{in}$ and $\tau _{out}$ are set as parameters indicative of spatial proximity within and between categories, respectively. Additionally, we use $class\_{num}$ to denote the number of categories.

\textbf{Match Within Categories:}
In this stage, we initiate by organizing the sequence of all images into a list denoted as $path\_List$. Our process starts with the first image in this list, for which we select two distinct templates, represented as $T_n(1,j,x',y')$, where $n=(1,2)$. The subsequent step involves systematically matching the defined template area with corresponding areas in all subsequent images within the $path\_List$, employing Eq. \ref{eq:10} for this purpose. If the match exceeds $\tau _{in}$, the two images are assigned to the same category.
We then move on to the second image. If it hasn't been matched yet, we select two templates, $T_n(2,j,x',y')$, and repeat the matching process. We use the variable $t$ to decide whether a template is matched or not during one matching process.
It's important to note that if the template doesn't find any matching images, the match is sought in the preceding images along the path, starting from the first image. Given that these images have already been categorized, the corresponding category number of the matched image is assigned to the template. We then go to the next image, repeating the above steps. This process continues until each image has been assigned its own category number, denoted as $c_j$. The algorithm is shown as Alg. \ref{alg:PFCFR1}. 

\begin{algorithm}
\caption{Match Within Categories}
\label{alg:PFCFR1}
\begin{algorithmic}[1]
\Require {All CFR images of M users, $\tau _{in}$, $\tau _{c}$, and the size of template $a,b$}
\Ensure $c_j$ for $j = 1,2,...M$
\Initialize{$class\_{num} = 0$ and set $c_j = \tau _c > M$ for $j = 1,2,...M$}
    \For{$i = 1,2,...,M$} \Comment{Match within categories}
    \State $t \gets 0$
    \If{$i=1$ or $c_i = \tau _c$}
      \State Choose $T_n(i,j,x',y')$ according to $a,b$
      \State $c_i \gets class\_{num}$
    \EndIf
        \For{$j = i+1,i+2,...,M$}
            \If{$c_j = \tau _c$}
                \State Compute $E_n(i,j)$
                \If{$E_n(i,j) >= \tau _{in}$}
                    \State $c_j \gets class\_{num}, t \gets 1$
                \EndIf
            \EndIf
        \EndFor
        \If{$t = 0$}
            \For{$k = 1,2,...,i$}
                \State Compute $E_n(i,k)$
                \If{$E_n(i,k) >= \tau _{in}$}
                    \State $t \gets 1$, $c_j \gets c_k$
                    \State Break the loop
                \EndIf
            \EndFor
        \EndIf
        \If{$t = 0$}
            \State $c_j \gets class\_{num}$
        \EndIf
        \State $class\_{num} \gets class\_{num}+1$
    \EndFor
\end{algorithmic}
\end{algorithm}

\textbf{Match Between Categories:}
The objective of this stage is to macroscopically merge similar categories. We utilize Eq. \ref{eq:10} to evaluate the similarity between the templates derived from the first-stage classification results. When the similarity measurement between any two templates exceeds $\tau _{out}$, we proceed to amalgamate all the CFR images associated with these templates into a single category. This merging process continues until there is no further change in the number of final classifications. The comprehensive algorithm for this stage is methodically outlined in Alg. \ref{alg:PFCFR2}.

\begin{algorithm}
\caption{Match Between Categories}
\label{alg:PFCFR2}
\begin{algorithmic}[1]
\Require {All templates and respective representative CFR images, $\tau _{out}$, $c_j$ and the size of template $a,b$}
\Ensure $c_j$ for $j = 1,2,...M$
    \State {Arrange $c_j$ in ascending order and update the index of the arranged $c_j$ to the new $c_j$ for $j = 1,2,...M$}
    \For{$i = 1,2,...,max(c_j)$} \Comment{Match between categories}
        \For{$j = 1,2,...,max(c_j)$}
            \If{$j \neq i$}
                \State Choose $T_n(i,j,x',y')$ according to $a,b$
                \State Compute $E_n(i,j)$
                \If{$E_n(i,k) >= \tau _{out}$}
                    \State Merge $c_i$ and $c_j$ into the same category
                \EndIf
            \EndIf
        \EndFor
    \EndFor
    \State {Arrange $c_j$ in ascending order and update the index of the arranged $c_j$ to the new $c_j$ for $j = 1,2,...M$}
\end{algorithmic}
\end{algorithm}

\subsubsection{ADCAM}
To assist in the pre-classification of ADCAM, we employ the iterative centroid-based clustering method to partition information such as the AOD, AOA, gains, and pathloss of each path into K clusters based on their distances from K class centers. 
To choose the appropriate value of K, we combine the Silhouette coefficients and Calinski-Harabasz methods to combine both the direct distances and covariance \cite{jain2010data}. This helps avoid scenarios where the graph of the clustering quality measure exhibits a smooth pattern (e.g., horizontal or continuously ascending/descending), making it difficult to ascertain the ideal K value.

\subsubsection{Data Fusing and Cleansing}
To ensure high-quality input data for our model, our initial step involves utilizing the two classification regions to refine the final classification categories. Subsequently, we apply data cleansing techniques to remove any anomalous fingerprints that might skew the analysis.
In the phase of combining classifications, we adopt a systematic approach. For instance, if there are only Class 0 and Class 5 ADCAM images found within the Class 0 CFR category, we assign values of [0, 0] and [0, 5] to the new categories as Class 0 and 1, respectively. This reclassification process is methodically applied across all categories.

During the data cleansing phase, we eliminate any category that contains a sample size of 2 or fewer. The rationale behind this decision is that such a small sample size in a category typically indicates an aberration, possibly due to multi-path features that significantly diverges from neighboring points. This could be attributed to erroneous recordings or the point being situated in a distinctly unique location. 

\subsection{Network Training}
\begin{figure}\vspace{-0mm}
	\begin{center}
		\centerline{\includegraphics[width=0.5\textwidth]{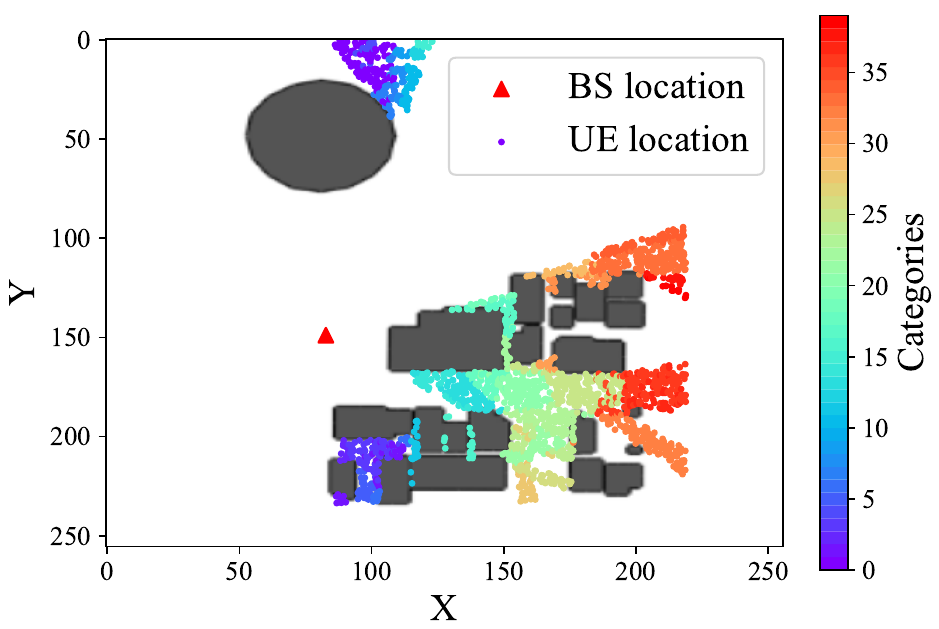}}  \vspace{-0mm}
		\caption{NLOS pre-classification.}
		\label{fig:visulizaion1} \vspace{-11mm}
	\end{center}
\end{figure}

After classifying the samples into different regions as shown in Fig. \ref{fig:visulizaion1}, we proceed with the network training process. This process consists of feature extraction, feature fusion, and filter regression, as illustrated in Fig. \ref{fig:framework}. The objective is to enable fingerprint localization, expressed as:
\begin{equation*}
    {\hat{\bf {p}}_{i,j}}={\mathbf{\mathcal G_{3,i}}}(\mathbf{\mathcal P}(\mathbf{{\mathcal G_{1}}}(H_{i,j}),{\mathbf{\mathcal G_{2}}(A_{i,j})}))
,\tag{9}\end{equation*}
where ${\mathbf{\mathcal G_{1}}}$ and ${\mathbf{\mathcal G_{2}}}$ represent the mappings built by deep residual network 18 (ResNet18 \cite{he2016deep}) from CFR or ADCAM images to a vector with a flattened length of $M_H$ or $M_A$ respectively. $\mathbf{\mathcal P}$ is a function that merges these two sets of features along the same dimensions. 
Finally, $\mathbf{\mathcal G_{3,i}}$ is a parallel multi-head mechanism in which the combined features are divided into different linear classifiers for regression operations. The variables $\sum_i i = I$ denote the number of final classifications, and each linear classifier corresponds to a specific region denoted as $i$. $\sum_i \sum_j i\times j = M$, and ${i,j}$ represents the $j$-th sample in the $i$-th class of the fingerprint after the heterogeneity matching. 

The loss function used for training is the mean-squared error (MSE) between the true positional coordinates $p$ and the predicted positional coordinates $\hat{p}$, calculated as:
\begin{equation*} \text {MSE Loss}=\frac {1}{n}\sum _{i=1}^{n}\left \|{\hat { \boldsymbol {p}}_{i}- \boldsymbol {p}_{i}}\right \|^{2},\tag{10}\end{equation*}
Here, $n = 16$ indicates the mini-batch size. Weight initialization is performed using Xavier initialization, and the Adam optimizer is utilized. The initial learning rate is set to 0.003.

\section{Numerical Results}\label{sec:intro}

This section demonstrates our framework's performance through simulations on scenario 00743 of the WAIR-D dataset. Our approach was trained for 150 epochs with the Adam optimizer with a learning rate of $3\times 10^{-3}$ in the Nvidia 3090.
Network architectures and hyperparameters were chosen via cross-validation for consistency across experiments unless stated otherwise.

\subsection{Datasets}
WAIR-D is a mmWave MIMO dataset for researchers to study and evaluate AI algorithms for wireless systems. It utilizes ray-tracing method, which is based on a high-frequency approximation to the Maxwell equations and describes the propagating field as a set of propagating rays, reflecting, diffracting, and scattering over environment elements to obtain the accurate characterization and simulation of electromagnetic propagation. 
The parameters used to generate the dataset are listed in Tab. \ref{tab:dataset}. 
\begin{table}[h!]
\normalsize
\centering
 \vspace{-2mm}
 \caption{Datasets parameters.}
 \begin{tabular}{c c | c c} 
 \toprule
 Parameters & Value & Parameters & Value\\ [0.2ex] 
 \midrule
 Carrier frequency & 60GHz & BS antenna & [1,64,1]\\ 
 Bandwidth & 0.05GHz & UE antenna & [1,1,1] \\
 Sub-carriers & 64 & Size of template & [8,16] \\
 \bottomrule
 \end{tabular}
 \label{tab:dataset}
 \vspace{-3mm}
\end{table}

\subsection{Visualization of Classification Areas}
Firstly, we visualize the classification results of the scene as shown in Fig. \ref{fig:visulizaion1}. It can be seen that our pre classification has excellent effectiveness, dividing an outdoor scene with a large area of $250\times250$ meters and many buildings into lots of small areas with irregular boundaries. The data within each region has relatively similar distribution characteristics, and there is good distinguishability between regions as well, greatly reducing the data distribution heterogeneity problem.
By leveraging these features, it can be used for channel estimation, channel inference, beamforming and so on.

\subsection{Effectiveness of Classification Methods}
We compare the proposed algorithm with the following baseline algorithms for NLOS when the antenna array is ULA. 
\begin{enumerate}
    \item Res\_CFR/Res\_ADCAM/Res\_CFRADCAM: Use CFR or ADCAM or CFR and ADCAM together as input for the feature extractor, and then put the obtained features into a linear classifier to obtain the predicted coordinates.
    \item Res\_multi\_CFRADCAM/Res\_multi\_CFRperfectADCAM: Compared to Res\_CFRADCAM, these two baselines put the obtained features into multiple linear classifiers to obtain the predicted coordinates. Res\_multi\_CFRperfectADCAM replaces ADCAM with AOA, AOD, distance, gain of the first arrival path as input. 
\end{enumerate}
\begin{figure}\vspace{-0mm}
	\begin{center}
		\centerline{\includegraphics[width=0.475\textwidth]{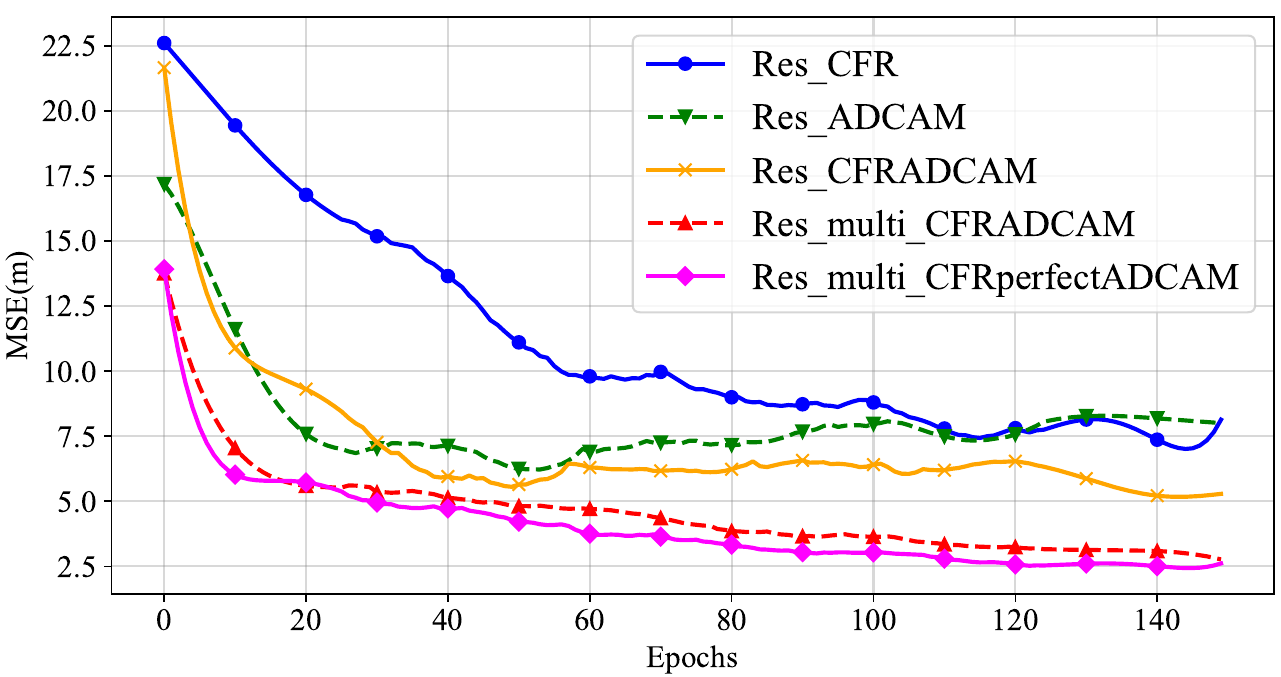}}  \vspace{-0mm}
		\caption{MSE with epochs of multiple processing results.}
		\label{fig:multiplemseepoch} \vspace{-8mm}
	\end{center}
\end{figure}
\begin{figure}\vspace{-0mm}
	\begin{center}
		\centerline{\includegraphics[width=0.475\textwidth]{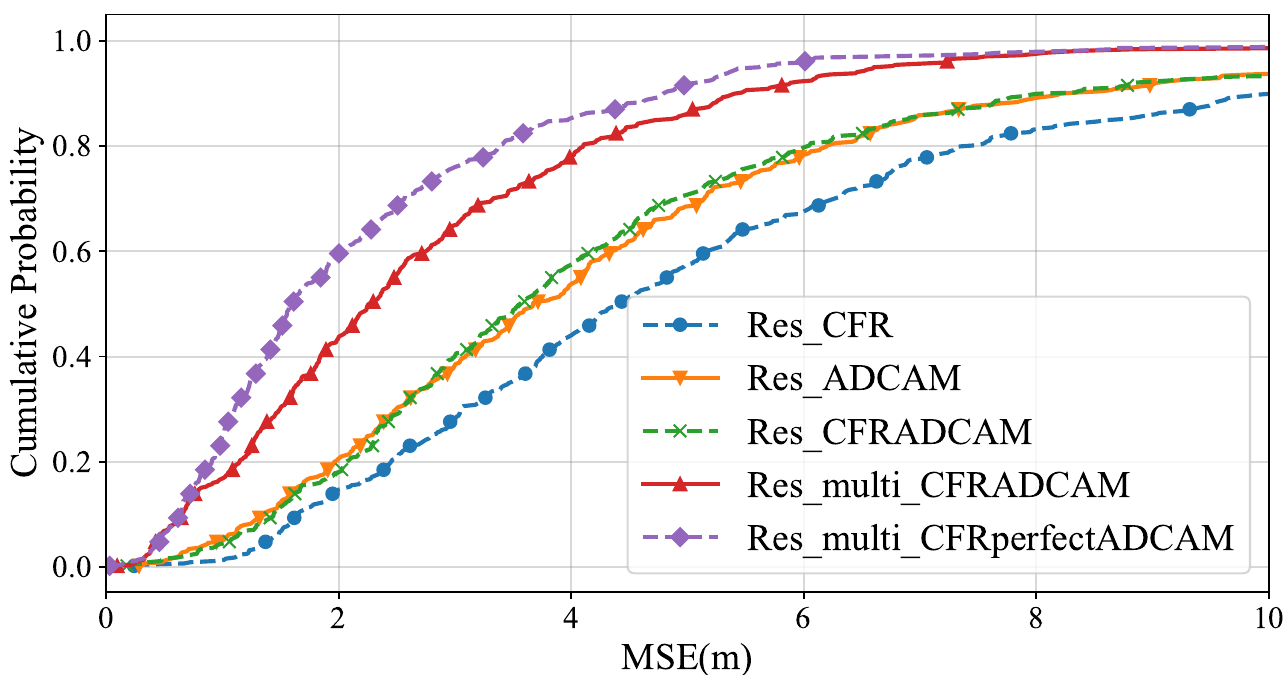}}  \vspace{-0mm}
		\caption{CDF curves of multiple processing results.}
		\label{fig:multiplecdf} \vspace{-12mm}
	\end{center}
\end{figure}

Fig. \ref{fig:multiplemseepoch} presents compelling evidence of the efficacy of our methodology. 
Firstly, as epochs increase, the MSE for Res\_ADCAM rapidly decreases, but ultimately close to that of Res\_CFR which is the highest. However, by adopting our AMDNLoc which fuses ADCAM and CFR deeply, the optimal MSE can be improved from 6.34 to 2.14 meters. 
This is because ADCAM is inherently sparser than CFR, and relying solely on either fingerprint can't fully represent the multi-path features.
What's more, AMDNLoc is more focused on the changes between the locations and the fingerprints with the similar data distribution, so it can better reflect the true nature of the location-fingerprint relationship.
Further supporting our findings, Fig. \ref{fig:multiplecdf} illustrates that over 60\% of positioning errors in our model fall within a 2-meter range. This is a significant increase compared to the baseline models, which only demonstrate a 20\% probability of positioning errors within the same range. These results unequivocally demonstrate the superiority of our integrated approach in achieving more accurate and reliable positioning in MIMO-OFDM systems.



\section{Conclusion}\label{sec:conclu}
In the paper, we proposed a novel multi-sources information fusion learning framework named AMDNLoc that makes full use of the multi-path feature across the frequency, power, angle and delay domain as fingerprints to tackle the inherent heterogeneity issue of fingerprint distributions. 
We explore a two-stage matched filter for PFCFR corresponding to the distribution of CFR, and fuse it with ADCAM classification region after centroid-based clustering method. 
What's more, a segment-specific linear classifier mechanism sharing the same feature extractor is utilized to build regression relationship between fingerprint and locations after eliminating the negative effect of the region covariant. 
Numerical experiments have shown that the AMDNLoc achieves SOTA results compared with traditional convolutional neural networks on
the WAIR-D.

\small
\bibliographystyle{IEEEtran}
\bibliography{bib}
\vspace{12pt}

\end{document}